# STIMULATED RADIATION COOLING *

E.G.Bessonov, R.M.Feshchenko
P. N. Lebedev Physical Institute RAS, Russia.

*Abstract*

In this paper the Stimulated radiation cooling (SRC) of ion beams and methods of formation of the broadband laser beams with given spectral distribution are discussed.

## INTRODUCTION

Beam cooling is the increase of six dimensional (6D) phase space density and reduction of the 6D emittance of the beam. The brightness $B$ of the particle beam (the current density per unit solid angle and per unit energy spread) is proportional to its 6D phase space density. Being an important parameter of the beam quality, it determines the luminosity of the colliding beams and brilliance of the light sources based on particle beams. Thus the cooling of the beams is of high importance for many applications. The rate of the beam density change in circular machines is determined by the 6D damping increment

$$\alpha_{6D} = \overline{div\ \vec{F}_{Fr}} = \frac{2}{\beta^2}\frac{\overline{P}_{Fr}(p)}{\varepsilon}|_s + \frac{\partial \overline{P}_{Fr}(p)}{\partial \varepsilon}|_s, \quad (1)$$

where $\vec{F}_{Fr} = -\alpha(\vec{r},p,t)\cdot \vec{n}_p$ is the friction force, $\alpha(\vec{r},p,t) = P_{Fr}(p,t)/c\beta$ is the frictional coefficient, $\vec{n}_p = \vec{p}/|\vec{p}|$ is the unit vector directed along the particle velocity, $\vec{p}$ is the particle momentum, $p = |\vec{p}|$, $\overline{P}_{Fr}(p)$ is the averaged rate of the particle energy loss due to friction, $\beta = v/c$ is the relative velocity of the particle, $\varepsilon$ is the energy of the particle, and index $s$ refer to the synchronous values. The equation (1) follows from the Robinson damping criterion [1], [2].

The value $\alpha_{6D} = 2\sum_{i}^{3}\alpha_i$ where $\alpha_i$ are the damping coefficients for the corresponding variables. From the equations of motion and for uncoupled vertical oscillations in a curvilinear coordinate system used in the particle accelerator theory the coefficients are

$$\alpha_x = \frac{1}{2}\left[\frac{\overline{P}_s}{\varepsilon_s} + \frac{\partial \overline{P}}{\partial \varepsilon}|_s - \frac{d\overline{P}}{d\varepsilon}|_s\right], \quad \alpha_z = \frac{1}{2}\frac{\overline{P}_s}{\varepsilon_s}, \quad \alpha_\varepsilon = \frac{1}{2}\frac{d\overline{P}}{d\varepsilon}|_s. \quad (2)$$

The energy loss function must increase with the energy for longitudinal cooling to occur. In the case of the synchrotron radiation cooling $\overline{P} = a_r\varepsilon^2$, $\partial \overline{P}/\partial \varepsilon = 2\overline{P}/\varepsilon$, $\alpha_\varepsilon = \overline{P}/\varepsilon$, $\alpha_{6D}|_{\beta=1} = 4\overline{P}/\varepsilon$.

Fast (enhanced) 6D cooling will occur if $\partial P/\partial\varepsilon|_s \gg \overline{P}/\varepsilon_s > 0$. One of the examples of the fast cooling is stimulated radiation ion cooling by broadband laser beams [3]. In this case the term "stimulated radiation cooling" (SRC) means a fast cooling of ion beam in the bucket by a broadband laser beam based on the backward Rayleigh scattering [4]. It does not refer to the stimulated emission. In [3] the nonlinear regime ($\overline{P}=0$ at $\varepsilon < \varepsilon_s$ and $\overline{P} > 0$ at $\varepsilon > \varepsilon_s$) was considered.

Note, that the beam density obeys exponential law only for the system described by linear differential equations. In this case all parts of the beam are cooled identically.

## FAST COOLING OF ION BEAMS

Below we will consider enhanced radiative ion cooling by a broadband laser beam interacting with electrons in the straight section of a storage ring with zero dispersion function [4]. The radial betatron and longitudinal planes are uncoupled, $\partial \overline{P}/\partial\varepsilon = d\overline{P}/d\varepsilon$ and the damping increments are

$$\alpha_x = \alpha_z = \frac{1}{2}\frac{\overline{P}_s}{\varepsilon_s}, \qquad \alpha_\varepsilon = \frac{1}{2}\frac{\partial \overline{P}}{\partial \varepsilon}|_s. \quad (3)$$

For the enhanced cooling $\alpha_\varepsilon \gg \alpha_x, \alpha_z$, which is the reason we consider cooling in the longitudinal plane only and determine the damping time energy variable $\tau_\varepsilon = 2/\alpha_\varepsilon$. Below linear and nonlinear versions of enhanced ion cooling will be considered.

### Linear version of the enhanced ion cooling.

In the linear version of cooling the friction power of ions is a linear function of the energy in the limits of the energy spread of the particle beam:

$$\overline{P} = \overline{P}_m \frac{\varepsilon - \varepsilon_c}{\varepsilon_s - \varepsilon_c + \sigma_\varepsilon} \text{ at } \varepsilon_c < \varepsilon < \varepsilon_s + \sigma_\varepsilon,$$

$$\overline{P} = 0 \text{ at } \quad \varepsilon < \varepsilon_c, \qquad \varepsilon > \varepsilon_s + \sigma_\varepsilon \quad (4)$$

where $\varepsilon_c < \varepsilon_s - \sigma_\varepsilon$, $2\sigma_\varepsilon$ is the energy spread of the beam. In this case, the minimum of the damping time occurs at the boundary of the linear regime corresponding to the energy $\varepsilon_c = \varepsilon_s - \sigma_\varepsilon$. According to (3), the damping time is $\tau_\varepsilon = 4\sigma_\varepsilon/\overline{P}_m$.

### Nonlinear version of the enhanced ion cooling.

In the nonlinear version of cooling the friction power of ions $\overline{P}$ can be complicated nonlinear function of the energy in the energy band of the beam being cooled. In the simplest case (stimulated radiation cooling)

$$\overline{P} = 2\overline{P}_m \frac{\varepsilon - \varepsilon_c}{\varepsilon_s - \varepsilon_c + \sigma_\varepsilon} \quad (\varepsilon_s < \varepsilon < \varepsilon_s + \sigma_\varepsilon),$$

$$\overline{P} = 0 \quad (\varepsilon < \varepsilon_s, \quad \varepsilon > \varepsilon_s + \sigma_\varepsilon), \quad (5)$$

where $\varepsilon_c = \varepsilon_s$. In this case the Robinson damping criterion in the form (1)-(3) does not work. It works separately at the energies $\varepsilon > \varepsilon_s$ (the rate of damping is 2 times less then in the previous case) and $\varepsilon < \varepsilon_s$ (no damping) leading to the average damping time equal to that in the linear case if average powers $\overline{P}_m$ in these cases are equal at the energy edges $\varepsilon = \varepsilon_s + \sigma_\varepsilon$.

Note that in the linear scheme the synchronous ions interact with the laser beam and emit scattered photons with power $\overline{P}_s = \overline{P}_m / 2$. It means that the ion beam is heated in this process due to the quantum nature of the light scattering. The equilibrium energy spread of the ion beam in this case is

$$\sigma_{\varepsilon,eq} = \hbar\overline{\omega}\sqrt{\overline{P}_s\tau_\varepsilon/\hbar\overline{\omega}} = \hbar\overline{\omega}\sqrt{2\sigma_\varepsilon/\hbar\overline{\omega}} \gg \hbar\overline{\omega}, \quad (6)$$

where $\hbar\overline{\omega} = \hbar\omega_l\gamma^2$ is the average energy of scattered photons, $\hbar\omega_l$ is the energy of laser photons, $\gamma = \varepsilon/m_ic^2$ and $m_ic^2$ are the relative energy and the rest energy of the ion.

In the nonlinear scheme ion energies tend to synchronous one if $\varepsilon > \varepsilon_s$ and do not change if $\varepsilon < \varepsilon_s$. Limiting energy spread in this case is about $\hbar\overline{\omega} = \hbar\omega_l\gamma^2 \ll \sigma_{\varepsilon,eq}$.

The nonlinear version of the laser cooling considered above is not optimal one. Faster increase of the power losses like

$$\overline{P} = \overline{P}_m \frac{\varepsilon - \varepsilon_s}{\varepsilon - \varepsilon_s + \sigma_c} \quad \text{at} \quad \varepsilon_s < \varepsilon < \varepsilon_s + \sigma_\varepsilon,$$

$$\overline{P} = 0 \quad \text{at} \quad \varepsilon < \varepsilon_s, \quad \varepsilon > \varepsilon_s + \sigma_\varepsilon, \quad (6)$$

and at $\sigma_c \ll \sigma_\varepsilon$ is preferable. Optimization of the nonlinear version is the topic for the future search.

Up to this moment we supposed that the laser beam is homogeneous in the limits of the ion beam. Ions both with positive and negative deviations from the synchronous orbit interact with the laser beam the same way. If the dispersion function of the storage ring differ from zero and the laser beam density decrease in the direction of the synchronous orbit then the emittance exchange between radial betatron and synchrotron oscillations will take place (wedge shape target) [5]. Fast cooling in the transverce plane will take place as well.

## METHODS OF FORMATION OF THE BROADBAND LASER BEAMS WITH GIVEN SPECTRAL DISTRIBUTION

The power $\overline{P}$ is determined by the spectral distribution of the laser intensity. The generation of a laser beam with a broad band frequency spectrum and sharp frequency cutoff is important problem in radiative cooling technique. Different schemes of generation can be used. 1) The necessary power can be generated by thermal optical sources or broadband lasers, filtered and then amplified by optical parametric amplifiers. 2) An undulator with a deflecting parameter $K \sim 1$ together with narrowband laser light can be used for the ion excitation in the interaction region (equivalent to the broadband laser light) [3]. In this case a small number periods tapered undulator with the magnetic field varying by definite law and monochromatic laser beam will be equivalent to laser with a broad band and a sharp frequency cutoff. 3) Successive frequency shift of a single mode laser by an acousto-optic modulator coupled to a passive ring cavity and other methods can also be used [6].

## REFERENCES

[1] K.W.Robinson, Radiation effects in Circular Electron accelerators, *Physical Review*, 1958, v.111, No 2, p.373-380; Electron radiation at 6 BeV, CEA Report No 14 (1956).
[2] E.G.Bessonov, The evolution of the phase space density of particle beams in external fields, arXiv:0808.2342v1; http://lanl.arxiv.org/abs/0808.2342.
[3] E.G.Bessonov, Some peculiarities of the Stimulated Radiation Ion Cooling, Bulletin of the American physical society, Vol.40, No 3, NY, May 1995, p.1196.
[4] E.G.Bessonov, Kwang-Je Kim, Radiative cooling of ion beams in storage rings by broad band lasers, Phys. Rev. Lett., 1996, v.76, No 3, p.431-434
[5] D.Neuffer, Principles and applications of muon cooling, *Particle accelerators*, v.14, 1983, p.75-90.
[6] S.N. Atutov1, R. Calabrese, V. Guidi et al., Generation of a frequency comb for white-light laser cooling of ions in a storage ring**,** Proc. 2006 European Particle accelerator Conference, EPAC2006, June 26-30 2006. Edinburgh, Scotland, http://cern.ch/AccelConf/e96/PAPERS/THPL/THP039L.PDF